\documentclass[aps,prl,reprint,tightenlines,superscriptaddress,amsmath,amssymb,showpacs,showkeys]{revtex4-1} 

\usepackage[utf8]{inputenc} 
\usepackage{graphicx} 
\usepackage{bm} 
\usepackage{natbib}
\usepackage{mathptmx}
\usepackage{epstopdf}
\usepackage{color}
\usepackage[english]{babel}
\usepackage{upgreek,xspace}
\usepackage[colorlinks=true,linkcolor=blue]{hyperref}

\newcommand\axi{\xi_{{\alpha}}}
\newcommand\axik{\xi_{{\alpha}_k}}
\newcommand\axikprime{\xi_{{\alpha}_{k'}}}
\newcommand\bxi{\xi_{\beta}}
\newcommand\bxik{\xi_{{\beta}_k}}
\newcommand\bxikprime{\xi_{{\beta}_{k'}}}
\definecolor{orangefonce}{rgb}{0.93,.46,.}
\definecolor{bleupale}{rgb}{0,0,.8}
\definecolor{greenfonce}{rgb}{0,.4,0}

\newcommand{\micron}{$\upmu$}

\begin{document}

\title{Light-mediated cascaded locking of multiple nano-optomechanical oscillators}
\author{E. Gil-Santos}
\affiliation{Mat\'{e}riaux et Ph\'{e}nom\`{e}nes Quantiques, Universit\'{e} Paris Diderot, CNRS UMR 7162, Sorbonne Paris Cit\'{e}, Paris, France}
\author{M. Labousse}
\affiliation{Mat\'{e}riaux et Ph\'{e}nom\`{e}nes Quantiques,  Universit\'{e} Paris Diderot, CNRS UMR 7162, Sorbonne Paris Cit\'{e}, Paris, France}
\author{C. Baker}
\affiliation{Mat\'{e}riaux et Ph\'{e}nom\`{e}nes Quantiques,  Universit\'{e} Paris Diderot, CNRS UMR 7162, Sorbonne Paris Cit\'{e}, Paris, France}
\author{A. Goetschy}
\affiliation{Mat\'{e}riaux et Ph\'{e}nom\`{e}nes Quantiques,  Universit\'{e} Paris Diderot, CNRS UMR 7162, Sorbonne Paris Cit\'{e}, Paris, France}
\affiliation{Institut Langevin, ESPCI Paristech, CNRS UMR 7587, PSL Research University,  1 rue Jussieu, 75005, Paris, France}
\author{W. Hease}
\affiliation{Mat\'{e}riaux et Ph\'{e}nom\`{e}nes Quantiques,  Universit\'{e} Paris Diderot, CNRS UMR 7162, Sorbonne Paris Cit\'{e}, Paris, France}
\author{C. Gomez}
\affiliation{Laboratoire de Photonique et de Nanostructures, CNRS UPR 20, Route de Nozay, 91460 Marcoussis, France}
\author{A. Lema\^{i}tre}
\affiliation{Laboratoire de Photonique et de Nanostructures, CNRS UPR 20, Route de Nozay, 91460 Marcoussis, France}
\author{G. Leo}
\affiliation{Mat\'{e}riaux et Ph\'{e}nom\`{e}nes Quantiques,  Universit\'{e} Paris Diderot, CNRS UMR 7162, Sorbonne Paris Cit\'{e}, Paris, France}
\author{C. Ciuti}
\affiliation{Mat\'{e}riaux et Ph\'{e}nom\`{e}nes Quantiques,  Universit\'{e} Paris Diderot, CNRS UMR 7162, Sorbonne Paris Cit\'{e}, Paris, France}
\author{I. Favero}
\affiliation{Mat\'{e}riaux et Ph\'{e}nom\`{e}nes Quantiques,  Universit\'{e} Paris Diderot, CNRS UMR 7162, Sorbonne Paris Cit\'{e}, Paris, France}

\date{\today} 
\begin{abstract}
Collective phenomena emerging from non-linear interactions between multiple oscillators, such as synchronization and frequency locking, find applications in a wide variety of fields. Optomechanical resonators, which are intrinsically non-linear, combine the scientific assets of mechanical devices with the possibility of long distance controlled interactions enabled by travelling light. Here we demonstrate light-mediated frequency locking of three distant nano-optomechanical oscillators positioned in a cascaded configuration. The oscillators, integrated on a chip along a coupling waveguide, are optically driven with a single laser and oscillate at gigahertz frequency. Despite an initial frequency disorder of hundreds of kilohertz, the guided light locks them all with a clear transition in the optical output. The experimental results are described by Langevin equations, paving the way to scalable cascaded optomechanical configurations.
\end{abstract}



\maketitle
Synchronization and frequency locking have been observed in a large variety of contexts ranging from physics to biology, {\it e.g.} in classical coupled pendula~\cite{Huygens1673}, in coupled lasers~\cite{Erneux1997}, and in the rhythmic beating of pacemaker cells~\cite{Sciences1975}. These phenomena have found practical applications in RF communication~\cite{Bregni2002}, signal-processing~\cite{Strogatz2003}, novel computing and memory concepts~\cite{Yamaguchi2008,Izhikevich2001}, clock synchronization and navigation~\cite{Bahder2009}, as well as in phased locked loop circuits~\cite{York1995}. For these applications, micro- and nano-mechanical devices are known to present opportunities of integration and scalability~\cite{Roukes2005,Rogers2004,Roukes2009,Seshia2013,Roukes2014,Seshia2014} but more recently, optomechanical systems further emerged as new appealing candidates. Indeed they support non-linearly coupled optical and mechanical modes~\cite{Aspelmeyer2014,Favero2009}, and add to the mechanics the assets of optical techniques in terms of precision and long-distance communications~\cite{Marquardt2011,Oehberg2012,Zhang2012,Shah2012,Tang2013,Milburn2012,Marquardt2013,Zhang2015}.  

Light injected in an optomechanical cavity can deform it under the action of optical forces, and in the dynamical back-action regime can amplify its mechanical motion. When amplification overcomes mechanical dissipation, the system transits to a stable limit cycle, often referred to as optomechanical self-oscillation~\cite{Karrai2004,Carmon2005}. In the last years, several studies investigated the synchronization of such optomechanical oscillators~\cite{Zhang2012,Shah2012,Tang2013,Zhang2015}. The synchronization of two oscillators placed close to contact and sharing a common optical mode was reported in~\cite{Zhang2012}. Recently, the same configuration was pushed up to 7 resonators~\cite{Zhang2015}. Two spatially-separated oscillators integrated in a common optical racetrack cavity were also synchronized in~\cite{Tang2013}. The possibility of locking two optomechanical systems without sharing a common optical mode was implemented as well in \cite{Shah2012}, in two steps and with two lasers. The optical output of a first laser-driven optomechanical oscillator was transduced into an electrical signal, which was carried away to fed a distant electro-optic modulator. The latter modulated a second laser driving a second optomechanical oscillator, ultimately insuring phase locking of the two. This master-slave electro-optic configuration enabled locking two oscillators at long distance but required one extra laser and modulator per added oscillator. An all-optical configuration linking multiple distant oscillators would allow for scalable optomechanical networks with controlled phase relations between sites. It remains to be explored.

\begin{figure}[!ht]
\begin{center}
\includegraphics[width=\columnwidth]{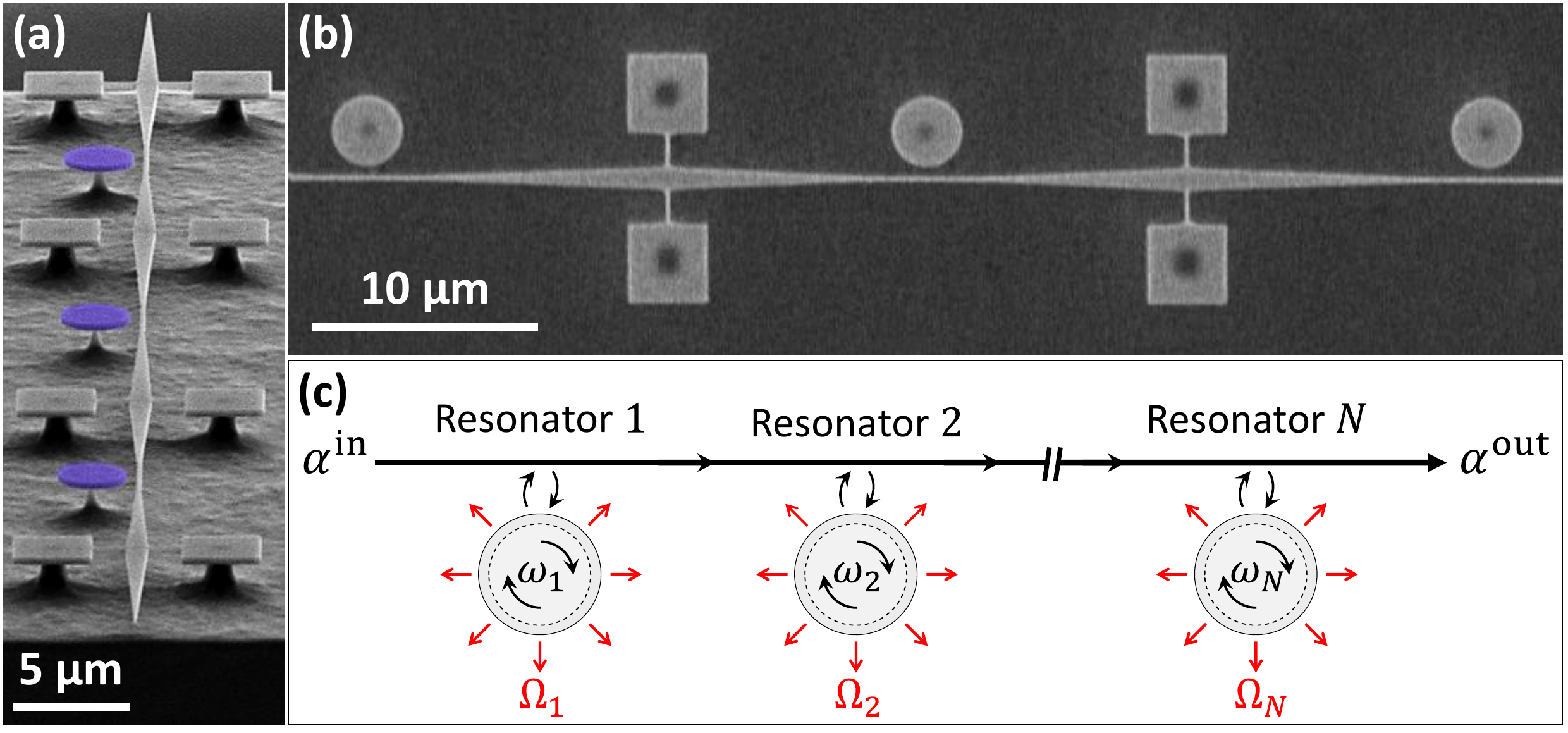}
\caption{(a) Side view and (b) top view, scanning electron microscope (SEM) images of a device consisting of 3 GaAs optomechanical disk resonators with identical nominal dimensions (radius of $1.5$ micron and thickness of $320$ nm) coupled to a common GaAs suspended tapered waveguide. The square-shaped pads are supports for the suspended optical waveguide and have no optomechanical impact. (c) A sketch of the unidirectional cascaded configuration containing multiple optomechanical disks along a common waveguide. }
\label{Fig1}
\end{center}
\end{figure}

Here, we demonstrate the all-optical light-mediated locking of multiple spatially distant optomechanical oscillators, achieved using a single laser source. Optomechanical disk resonators, each supporting its own localized optical and mechanical mode, are placed in a cascaded configuration and unidirectionally coupled trough a common optical waveguide [See Fig.~\ref{Fig1}]. Optical and mechanical modes show a remarkably low site-to-site disorder of a few $10^{-4}$, which results from moderate residual fabrication imperfections. We take advantage of thermo-optic effects to lower even further the optical disorder (below $10^{-5}$) and inject light simultaneously in all the resonators, eventually locking their very high-frequency (GHz) mechanical oscillations. We present results for configurations with one, two and three optomechanical oscillators. The experimental results are in qualitative agreement with theoretical calculations based on a minimal model that considers a single optical and mechanical mode per disk resonator. The results for optical and mechanical fields are obtained by solving the corresponding stochastic Langevin equations within the truncated Wigner approximation~\cite{Wigner2005,RMP2013}, an approach that can be extended to arbitrary complex cascaded configurations. Optomechanical cascades are naturally scalable and lend themselves to both on-chip integration, like demonstrated here, or long-distance fibered networks. 

On the device side, we employ gallium arsenide (GaAs) optomechanical disk resonators. Arrays of miniature disks supported on their pedestal as shown in Fig. ~\ref{Fig1}(a) and (b), are fabricated out of a GaAs/Aluminium Gallium Arsenide (AlGaAs) hetero-structure wafer~\cite{Ding2015}. A picture of the complete sample structure is shown in the Supplemental Material (Fig. S1). Miniature disks support both optical whispering gallery modes (WGMs) and mechanical radial breathing modes (RBMs)~\cite{Favero2010} that strongly couple through radiation pressure and photoelasticity, reaching an optomechanical coupling $g_0$ in the MHz range~\cite{Favero2014}. This enables fine optical control of mechanical motion in a large variety of physical environments, from cryogenic quantum operation to liquid immersion~\cite{Favero2013,Favero2015}. Monochromatic light at $\lambda =1.3$ \micron m is evanescently coupled into the disks through integrated GaAs tapered waveguides \cite{baker2011guides}, which also embed inverted tapers as input and output ports to suppress light back-reflections (see Supplemental Material Fig. S2). The disks dimensions are 320 nm in thickness and 1.5 to 2 \micron m in radius. The pedestals are 1.7 \micron m high and their radii are smaller than 150 nm to minimize anchoring mechanical losses~\cite{Favero2013}. On the same chip, we include waveguides that address one, two and three disks in a unidirectional fashion. Disks are spatially separated by approximately 25 \micron m thereby avoiding direct mechanical and near-field optical couplings.\\

On the theoretical side, we model the driven-dissipative dynamics of $N$ sequentially coupled optomechanical resonators (see sketch in Fig.~\ref{Fig1}(c)) by applying a stochastic method based on the truncated Wigner representation \cite{Wigner2005,RMP2013,JuanPHD}. Note that this is equivalent to the Langevin approach in Ref. \cite{Marquardt2013}. In the truncated Wigner approximation, the master equation for the density matrix is mapped onto a differential equation for the Wigner function, which is truncated
by retaining only its lowest order derivatives. The resulting Fokker-Planck equation is converted into stochastic Ito differential equations for the $2 N$ complex scalar fields $\left\lbrace \alpha_k\right\rbrace_{k=1,\ldots,N}$ for the optics and $\left\lbrace\beta_k\right\rbrace_{k=1,\ldots,N}$ for the mechanics. This approach is valid for moderate nonlinearities for which the Wigner function remains positive (hence neglecting photon and/or phonon blockade effects~\cite{Juan2014}), which is the case in the present context. In the driving rotating frame, the obtained equations read:
\begin{equation}
\left\{
    \begin{array}{ll} 
    \dot{\alpha}_k=\left(i\left(\Delta_k+g_0(\beta_k+\beta_k^*)\right)-\dfrac{\kappa}{2}\right)\alpha_k+i\sqrt{\kappa^e_k}\alpha^{\mathrm{in}}_k+\axik\\
     \dot{\beta}_k=-\left(i\Omega_k+\dfrac{\Gamma}{2}\right)\beta_k+ig_0\left(\alpha_k\alpha_k^*-\dfrac{1}{2}\right)+\bxik,
     \end{array}
\right.
\label{Equationoptomec_semiclassic}
\end{equation}
similar to Ref.~\cite{Li2016}, where $\Omega_k$ denotes the mechanical frequency of the $k$-th optomechanical resonator, while $\Delta_k = \omega_k - \omega_p$ is the optical detuning between the $k$-th cavity eigenfrequency and the optical drive. The optical and mechanical loss rates are $\kappa$ and $\Gamma$ respectively. The stochastic terms $\axi$ ($\bxi$) denote the optical (mechanical) Langevin noise at room temperature, with correlations given by $\left\langle \axik(t)\axikprime(t')\right\rangle = \dfrac{\kappa}{2}\delta(t-t')\delta(k-k')$ and $\left\langle \bxik(t)\bxikprime(t')\right\rangle = \dfrac{\Gamma}{2}\left(n_{\mathrm{th}}+\dfrac{1}{2} \right)\delta(t-t')\delta(k-k')$. The optical input $\alpha_{k}^{\mathrm{in}}=\alpha_k^{\mathrm{\mathrm{pump}}}+\alpha_k^{\mathrm{\mathrm{cav}}}$ is the sum of two distinct coherent contributions.  $\alpha_k^{\mathrm{\mathrm{pump}}}=\sqrt{P_{\mathrm{laser}}/(\hbar \Omega_{\mathrm{L}})}e^{i \phi_k}$ is the driving input travelling up to the $k$-th cavity, while  $\alpha_k^{\mathrm{\mathrm{cav}}}=-\sum_{\ell<k}\sqrt{\kappa_{\ell}^e}\alpha_{\ell}e^{i\phi_{\ell\rightarrow k}}$ is the input field leaking out of the disks upstream. \\

\begin{figure}[!ht]
\begin{center}
\includegraphics[width=1\columnwidth]{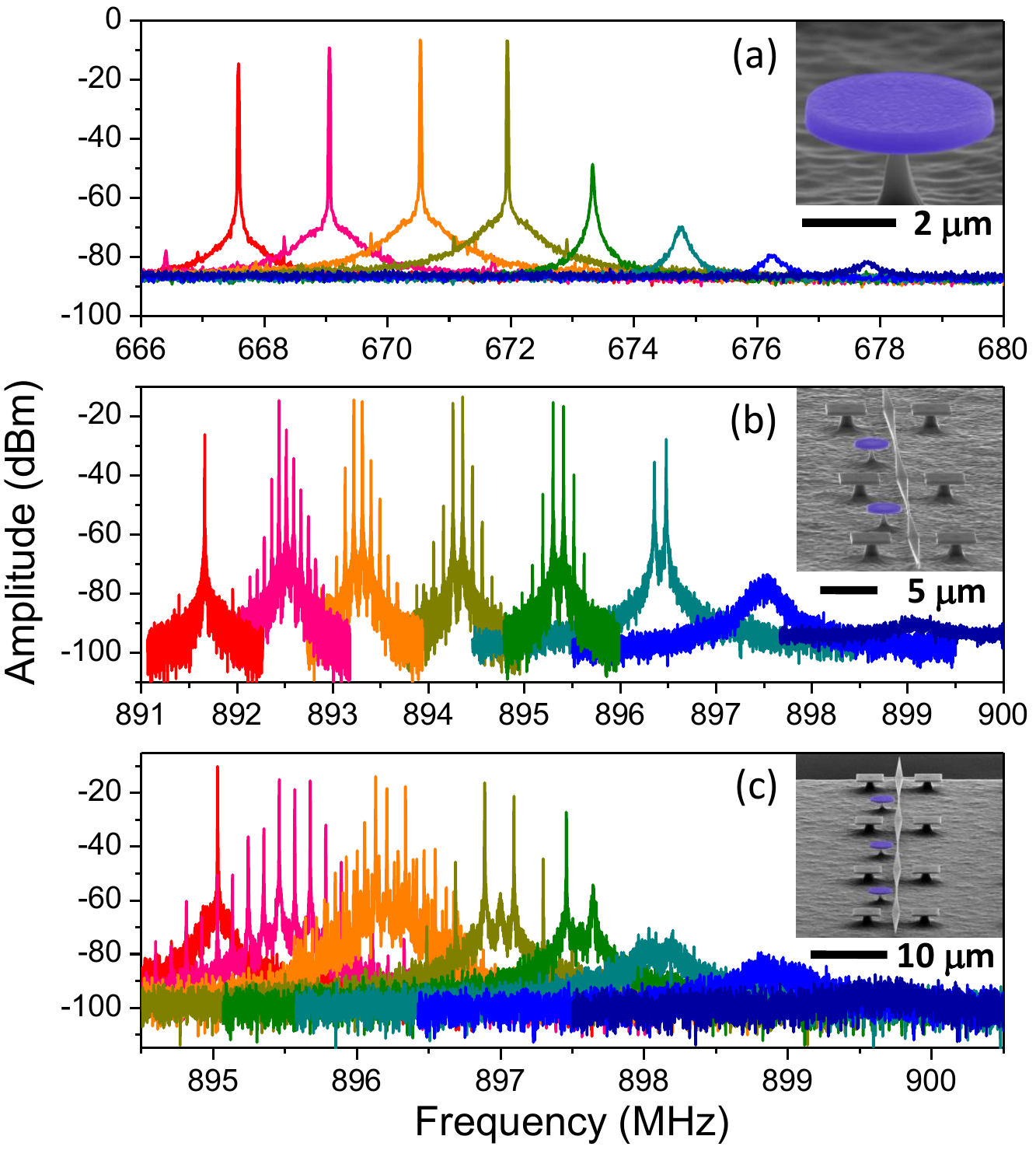}
\caption{a) From right (blue curve) to left (red curve), power spectral densities of a single optomechanical oscillator ($2$~\micron m radius and $320$ nm thickness) for increasing laser wavelength $\lambda$ (Supplemental Material) and $P_{\mathrm{laser}}=1$ mW. We first observe the background thermal motion, then the motion is amplified and finally reaches the self-oscillation regime. b) From right (blue) to left (red), power spectral density of two optomechanical oscillators ($1.5$~\micron m radius and $320$ nm thickness) for increasing laser wavelength (see Supplemental Material). We successively observe the thermal vibration of both resonators, then their amplification, the self-oscillation, the nonlinear mixing of their resonances, and finally their phase-locking. c) Same as in b), but with $3$ optomechanical oscillators.}
\label{Fig2}
\end{center}
\end{figure}

Experimental and theoretical methodologies being introduced, let us now discuss the core of our results. We experiment first on a waveguide evanescently coupled to a single disk. By lowering the laser input power $P_{\mathrm{laser}}$ down to 1 $\mu$W, and scanning the laser wavelength, we measure the optical spectrum of the system through the waveguide transmission. We observe WGMs resonances (Supplemental Material Fig. S3), in the form of doublets corresponding to hybrid clockwise/counter-clockwise modes ~\cite{DingSPIE2010}, with intrinsic optical quality factor of about $10^5$. Then, we increase $P_{\mathrm{laser}}$ to 1 mW and observe a thermal distortion of the optical resonance while scanning (Supplemental Material Fig. S3). This is known to result from the thermo-optic effect \cite{Ding2011} induced by the laser heating of the GaAs disk \cite{parrain_origin_2015}. The temperature rise in the disk $\Delta T$ can be estimated from the shift of the optical WGM resonance $\Delta \lambda$ using the relation $\Delta T=n(\Delta \lambda/\lambda)/(dn/dT)$, with $n$ the disk refractive index. At the largest optical power employed in our experiments, $\Delta \lambda \simeq 5$ nm, corresponding to $\Delta T\simeq 60$K. In this large optical power regime, we now step-scan the laser wavelength over the blue flank of the optical WGM resonance, and acquire the radio-frequency spectrum of transmitted light for each wavelength step. The results are shown in Fig.~\ref{Fig2}(a). When sufficiently blue-detuned (blue curve), the power in the cavity is too low to induce optomechanical dynamical back-action and the spectrum reveals the sole thermomechanical disk vibration. Measurements in this regime show an intrinsic mechanical quality above $10^3$. As the laser wavelength, and consequently the optical power in the cavity, are progressively increased, the mechanical motion gets amplified trough dynamical back-action (blue to green curves in Fig.~\ref{Fig2}(a)), until the optomechanical amplification and nonlinearity finally attract the resonator to the self-oscillation limit cycle associated to an abrupt line-narrowing in the spectrum (green to red curves in Fig.~\ref{Fig2}(a)). At this stage, we obtain a laser-driven nano-optomechanical oscillator, whose optical output will be propagated to the following disks. Note that the mechanical properties of GaAs disk are also affected by the laser-induced temperature increase, mainly because of the material softening, and a drift towards lower mechanical frequencies is observed as the optical detuning $\Delta$ is progressively decreased (blue to red curve, see Supplemental Material). 

We then experiment on a waveguide addressing two cascaded disks. When using low enough laser power, we observe a pair of doublets corresponding to the resonances of each disk (Supplemental Material Fig. S4(b)). Doublets are separated by about $400$ pm in wavelength, showing a relative optical disorder of $\sim 3.10^{-4}$. We then set the laser power to 1 mW, inducing thermal distortion of the optical resonances. The related thermo-optic wavelength shift is much larger than the initial optical disorder, enabling simultaneous injection of light in both disks with the same single laser (see Supplemental Material). Such thermo-optic compensation enables us to reduce the effective optical disorder at a value that we estimate below $10^{-5}$ (see Supplemental Material). Figure~\ref{Fig2}(b) shows the evolution of the RF spectrum of light outcoupled from the waveguide when the laser wavelength progressively increases starting from the blue flank of the WGMs resonances. Again, when the effective power in the disk is sufficiently low, we measure the mere thermomechanical vibration of both disks. As a consequence of the low mechanical disorder, smaller than the mechanical linewidth, the independent mechanical resonances of the two disks are spectrally overlapped and indistinguishable (blue curve in Figure~\ref{Fig2}(b)). As the laser wavelength further increases, therefore increasing the effective power in the disks, the motion of both disks is amplified. At this stage, thanks to this line narrowing, their respective resonances become distinguishable (light blue trace), revealing a mechanical disorder of about $10^{-4}$. Further increasing the laser wavelength, the two disks enter into the self-oscillation regime, and multiple resonance peaks arise. This set of peaks at frequencies $\lbrace\omega=\Omega_1 \pm p(\Omega_1-\Omega_2)\rbrace_{p\in \mathbb{N}}$ are generated by nonlinear frequency mixing. Ultimately, when the power in the two disk is large enough, the multiple peaks merge into a single one, indicating frequency locking of the two oscillators.\\
\\
\begin{figure}[!ht]
\begin{center}
\includegraphics[width=1.0\columnwidth]{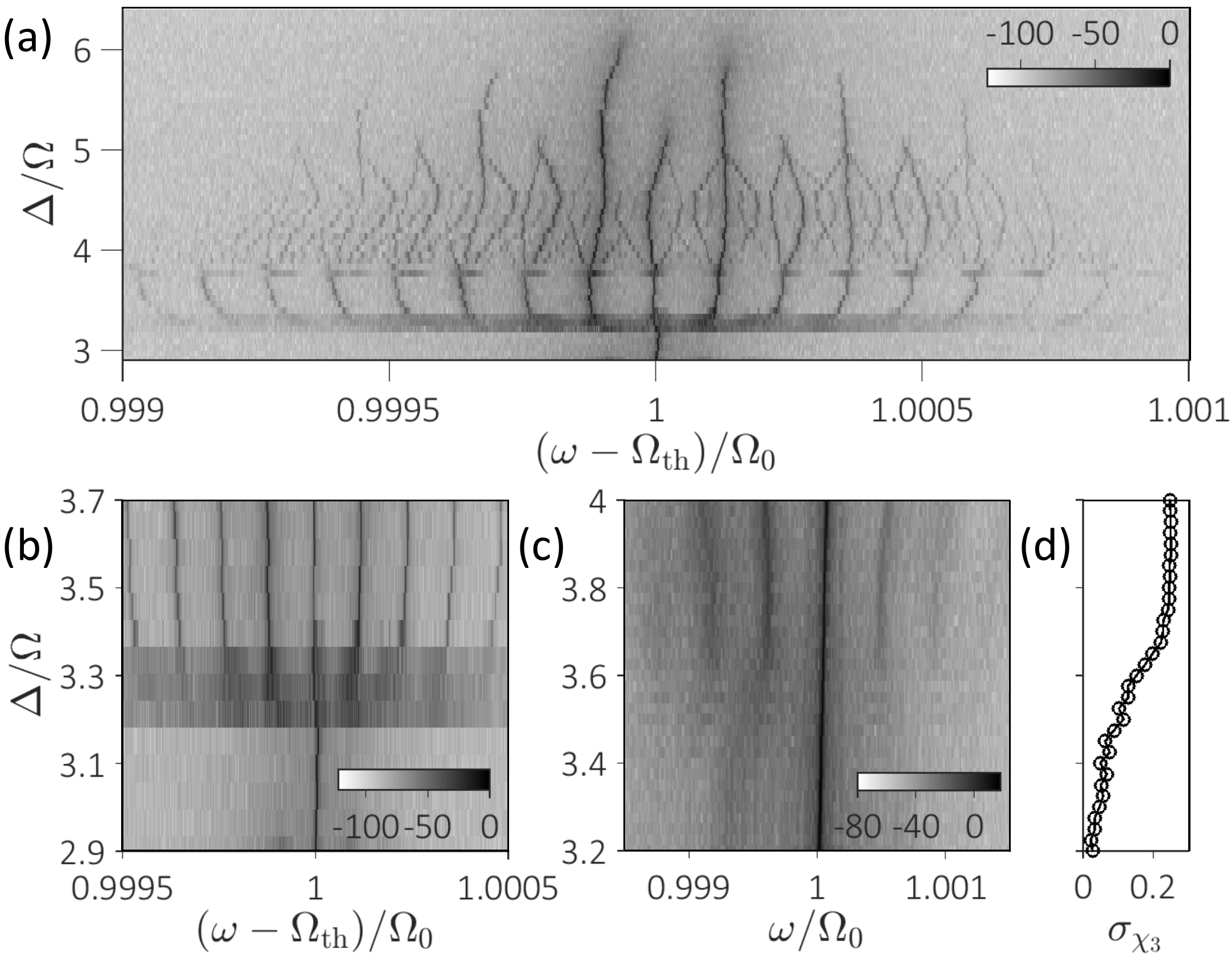}
\caption{(a) Experimental results: gray-scale map of the power spectral density for the system of $3$ optomechanical oscillators as a function of the normalized detuning $\Delta/\Omega$. Note that the thermo-mechanical shift has been subtracted for clarity, $\Omega_{th}$ being the thermally shifted mechanical frequency. (b) Zoom in (a). (c) and (d) Numerical results obtained with the model described in the text. (c) Gray-scale map of the power spectral density for $3$ oscillators as a function of the detuning $\Delta/\Omega$ for a mechanical disorder $\epsilon=5.10^{-4}$. (d) Standard deviation $\sigma_{\chi_3}$ of the synchronization order parameter as a function of the detuning $\Delta/\Omega$ (same vertical axis as in (c)), transiting to zero as phase locking occurs.}
\label{Fig3}
\end{center}
\end{figure}
Now we consider the case of three optomechanical resonators and follow the very same procedure, which is intrinsically scalable. At low power, we observe three doublets corresponding to the resonances of each disk (Supplemental Material Fig. S4(c)). The three disks show the same level of relative optical disorder as the two resonators above. Here again, at high power, the initial optical disorder is corrected by means of the thermo-optic compensation (Supplemental Material Fig. S4). Figure~\ref{Fig2}(c) shows a similar evolution of the RF mechanical spectrum when increasing the optical wavelength (blue to red trace). When the the detuning is large, the thermal vibrations of the disks are measured and the independent resonances are indistinguishable. As the laser wavelength increases, the respective mechanical resonances of the three disks become distinguishable thanks to optomechanical amplification and line-narrowing (blue to green trace). Further increasing the laser wavelength, the three disks enter into the self-oscillation regime, and multiple resonance peaks arise due to nonlinear frequency mixing (green to orange trace). Then, two of the oscillators lock as the density of peaks diminishes (pink trace). Ultimately, the multiple peaks merge into a single one, indicating the frequency locking of the three oscillators (red trace). More details will become apparent when analyzing this locked cascaded configuration in light of our model. 

The full spectrum evolution as a function of laser wavelength is shown in Fig.~\ref{Fig3}(a) for the three oscillators case. In this latter figure, we have subtracted for clarity the mechanical frequency drift induced by thermal softening associated to laser heating. At small wavelength (large detuning), the spectrum shows a few dark lines corresponding to the onset of self-oscillation for the three resonators. The density of lines increases with the laser wavelength due to non-linear mixing, before the spectrum simplifies as two of the three oscillators get locked together at $\Delta/\Omega\sim$3.8. The final step is the locking of the three oscillators, with a marked transition to a single peak (see a zoom in Fig.~\ref{Fig3}(b)). 

In Fig.~\ref{Fig3}(c), we compare the experimental results with the phenomenology predicted by the theoretical model. In order to account for the mechanical frequency disorder, we consider as example $\Omega_1=\Omega(1 -\epsilon)$, $\Omega_2=\Omega$, $\Omega_3=\Omega(1+1.23\epsilon)$, with $\epsilon=5.10^{-4}$. For simplicity, apart from the mechanical frequency disorder, we consider identical parameters for the three optomechanical resonators, namely  $\kappa/\Omega=7.4$, $\kappa_{\ell}^e= \kappa^e=2.3 \Omega$, $\Gamma/\Omega=1.3 \times 10^{-3}$. Thanks to the thermo-optical compensation (see discussion above and Supplemental Material), the optical detuning is set to be identical for all optical resonators ($\Delta_k=\Delta$). We also set $g_0/\Omega=5.10^{-4}$ for the optomechanical coupling and $P_{\mathrm{laser}}=800$ \micron W, following experimental conditions. The evolution of the noise power density with the detuning is plotted in Fig.~\ref{Fig3}(c) and shows that the phase-locked state is achieved after an intermediate regime displaying nonlinear frequency mixing, like observed in experiments. A complementary view of the transition is reported in Fig.~\ref{Fig3}(d):  the standard deviation $\sigma_{\chi_{3}}$ of the mechanical phase order parameter $\chi_3 = \tfrac{1}{3} \left \vert \sum_{j=1}^3  \tfrac{\beta_j}{\vert\beta_j \vert} \right \vert $ is shown versus normalized detuning. Locking of the oscillators is obtained for $\sigma_{\chi_{3}} = 0$. For lower mechanical disorder, our simple model predicts that phase-locking occurs without an intermediate nonlinear frequency mixing, therefore more directly than experimentally. Such a difference originates in part from the simplicity of our calculations where the optomechanical coupling and quality factor have been set equal for all resonators. Additional simulations (not shown) indeed indicate that individual self-oscillation thresholds are very sensitive to disorder in the optomechanical coupling parameter. In spite of these simplifications, our numerical results based on a minimal model describe experiments well and capture the right order of magnitude for the relevant quantities.

In conclusion, we have experimentally demonstrated locking of two and three very-high frequency optomechanical oscillators in a cascaded optical configuration. We have employed a single laser and exploited the optical coupling between distant mechanical systems mediated by an integrated waveguide. With long-distance optical interactions, our results pave the way to large-scale cascades and networks of optomechanical oscillators with controlled phase relations. Our semiconductor devices being integrable, a variety of on-chip configurations can be envisioned as well, for example in two dimensions for the development of topological optomechanics ~\cite{Hafezi,MarquardtTopology} or for new devices in signal processing and sensing. More generally, the study of nonlinear dynamical systems connected by light will offer an unprecedented degree of control of interactions, be it in the classical or quantum regime. 
 
\begin{acknowledgments}
We acknowledge fruitful exchanges with J. Restrepo,  F. Storme and W. Hease and the support of the ERC via the GANOMS project. 
\end{acknowledgments}

\bibliographystyle{apsrev}
\bibliography{Lockingbiblio}

\end{document}